\documentclass[prl,twocolumn,showpacs,preprintnumbers,amsmath,amssymb]{revtex4}

\usepackage{graphicx}
\usepackage{dcolumn}
\usepackage{bbm}

\begin{document}

\title{Unattainability of a purely topological criterion for the existence of a 
phase transition for non-confining potentials}

\author{Michael Kastner}
\email{Michael.Kastner@uni-bayreuth.de}
\affiliation{Physikalisches Institut, Lehrstuhl f\"ur Theoretische Physik I, Universit\"at 
Bayreuth, 95440 Bayreuth, Germany}

\date{April 29, 2004}

\begin{abstract}
The relation between thermodynamic phase transitions in classical systems and topology 
changes in their configuration space is discussed for a one-dimensional, analytically 
tractable solid-on-solid model. The topology of a certain family of submanifolds of 
configuration space is investigated, corroborating the hypothesis that, in general, a change 
of the topology within this family is a necessary condition in order to observe a phase 
transition. Considering two slightly differing versions of this solid-on-solid model, one 
showing a phase transition in the thermodynamic limit, the other not, we find that the 
difference in the ``quality'' or ``strength'' of this topology change appears to be 
insignificant. This example indicates the unattainability of a condition of exclusively 
topological nature which is sufficient as to guarantee the occurrence of a phase transition 
in systems with non-confining potentials.
\end{abstract}

\pacs{05.70.Fh, 02.40.-k, 68.35.Md}

\maketitle

Phase transitions, like the boiling and evaporating of water at a certain temperature and 
pressure, are common phenomena both in everyday life and in almost any branch of physics. 
Loosely speaking, a phase transition brings about a sudden change of the macroscopic 
properties of a system while smoothly varying a parameter (the temperature or the pressure 
in the above example). The mathematical description of phase transitions is conventionally 
based either on Gibbs measures on phase space or on (grand)canonical thermodynamic 
functions, relating their loss of analyticity to the occurrence of a phase transition. Such 
a nonanalytic behavior can occur only in the thermodynamic limit \cite{YangLee,Georgii}, in 
which the number of degrees of freedom $N$ of the system goes to infinity. Conceptually, the 
necessity of the thermodynamic limit is an objectionable feature: firstly, the number of 
degrees of freedom in real systems, although possibly large, is finite, and, secondly, for 
systems with long-range interactions, the thermodynamic limit may even not be well defined.

Recently, an alternative approach to phase transitions has been proposed, which connects the 
occurrence of a phase transition to certain properties of the potential energy $V$, 
resorting to {\em topological}\/ concepts. The conceptual advantages of this topological 
approach are twofold: The microscopic Hamiltonian dynamics, which is at the basis of the 
thermodynamic behavior of the system, can be linked via the Lyapunov exponents to the 
topological structure considered \cite{CaPeCo1}, thus rendering the topological approach a 
very fundamental one. Furthermore, in contrast to the conventional approach, there exists a 
``natural'' generalization of the concept of phase transitions to finite systems.

The topological approach is based on the hypothesis \cite{CaCaClePe} that phase transitions 
are related to topology changes of submanifolds $M_v$ of the configuration space of the 
system, where the $M_v$ consist of all points $q$ of the configuration space for which 
$V(q)/N \leqslant v$, i.\,e., their potential energy per degree of freedom is equal to or 
below a certain level $v$. This hypothesis has been corroborated by numerical as well as 
exact analytical results for some model systems showing first-order \cite{Angelani_etal} as 
well as second-order \cite{CaPeCo1,CaCoPe,CaPeCo2,GriMo} phase transitions. A major 
achievement in the field is the 
recent proof of a theorem, stating, loosely speaking, that, for systems described by smooth, 
finite-range, and confining potentials, a topology change of the submanifolds $M_v$ is a 
{\em necessary}\/ criterion for a phase transition to take place \cite{FraPeSpi}.

Albeit necessary, such a topology change is clearly not {\em sufficient}\/ to entail a phase 
transition. This follows for example from the analytical computation of topological 
invariants in the XY model \cite{CaCoPe,CaPeCo2}, where the number of topology changes 
occurring is shown to be of order $N$, but only a single phase transition takes place. So 
topology changes appear to be rather common, and only particular ones are related to phase 
transitions. While some ideas relating the ``strength'' of a topology change to the 
occurrence of a phase transitions have been put forward \cite{CaPeCo2}, a sufficient 
criterion on the quality of a topology change is still lacking, and the quest to seek for 
one can be considered as {\em the}\/ fundamental open problem in the field.

The objective of the present Letter is to shed some light on how such a sufficient criterion 
might (not) look like. To this purpose, a so-called solid-on-solid model proposed by 
Burkhardt \cite{Burkhardt} is considered. This model has the---for our purposes 
desirable---features to be (i) analytically solvable, and (ii) sensitive upon a slight 
modification of the model, in the sense that in one case it {\em does}\/ exhibit a phase 
transition, whereas in the other case {\em not}. By investigating analytically the 
topological properties of the submanifolds $M_v$ of the configuration space, we find that, 
in both these cases, a topology change takes place. In the case of the model exhibiting a 
phase transition, this transition is related to the topology change in accordance with the 
above mentioned topological hypothesis. Comparing, however, this topology change to the one 
in the model without a phase transition, a significant difference in the ``strength'' of the 
topology does {\em not}\/ appear to be present. Hence, the results presented in this Letter 
indicate that a discrimination between topology changes which entail a phase transition, and 
those which don't, may in general not be possible on topological grounds exclusively. This 
result puts the search for a sufficiency criterion on the topology change into a completely 
new perspective.

{\em Burkhardt model:} We consider a one-dimensional mod\-el on a lattice with real, 
continuously varying variables $q_i$. The Hamiltonian function is given by
\begin{equation}\label{Hamiltonian}
{\mathcal H}(q)=\sum_{i=1}^N \big[ \left| q_{i+1}-q_i \right| + U(q_i) \big],
\end{equation}
where $q=(q_1,...,q_N)$ is a state of the system. Periodic boundary conditions 
$q_{N+1}\equiv q_1$ are assumed. The so-called pinning potential $U$ is a real valued 
function, bounded below and above, with a unique infimum at zero. This one-dimensional 
system was introduced in \cite{Burkhardt} to model the localization-delocalization 
transition of an interface in a two-dimensional system. The pinning potential tends to 
localize the ``interface'' (i.\,e., the values of the $q_i$) around zero. The above 
Hamiltonian describes a {\em static}\/ model without a kinetic term, but it can be extended 
straightforwardly to include the dynamics as well.

The thermodynamic behavior of this system can be analyzed analytically by rewriting the 
partition function in terms of an integral transfer operator. Then, the eigenvalue equation 
of this operator can be transformed into a one-dimensional Schr\"odinger type equation. In 
doing so, the problem of finding a localization-delocalization transition is mapped onto the 
question whether there exist bound state solutions of the Schr\"odinger equation for certain 
potentials. For the example of a square well pinning potential
\begin{equation}\label{squarewell}
U(x)=\left\{
\begin{array}{r@{\quad\mbox{for}\quad|x|}c@{1,}}
-1& \leqslant\\
0 & >
\end{array}
\right.
\end{equation}
the latter problem is analyzed explicitly in \cite{Burkhardt}. It is found that the 
existence of a phase transition depends on the domain of the Hamiltonian function 
(\ref{Hamiltonian}):
\renewcommand{\labelenumi}{(\alph{enumi})}
\begin{enumerate}
\item For the $q_i$ taking on values on the semi-infinite line, $q_i \in 
[\,0,+\infty)={\mathbbm R}_0^+$, a second-order localization-delocalization transition is 
observed.
\item In case of the $q_i$ having values from the real numbers, $q_i \in 
(-\infty,+\infty)={\mathbbm R}$, no transition takes place.
\end{enumerate}
It is argued in \cite{Burkhardt} that this result generalizes to a large class of pinning 
potentials $U$. Note that the existence of a phase transition in such one-dimensional 
systems is by no means contradictory to van Hove's theorem: the conditions assumed in van 
Hove's work \cite{vanHove} simply don't apply to the Burkhardt model (\ref{Hamiltonian}), 
neither do they to many other models. For a more general theorem and further interesting 
aspects of phase transitions in one-dimensional systems see \cite{CuSa}.

{\em Topological approach:} Certain submanifolds $M_v$ of the configuration space are taken 
as a starting point for the topological approach. Due to the absence of a kinetic term in 
(\ref{Hamiltonian}), the configuration space $\Gamma$ is identical to the domain of the 
Hamiltonian. For a system consisting of $N$ degrees of freedom, we have 
$\Gamma_a=\left({\mathbbm R}_0^+\right)^N$ and $\Gamma_b={\mathbbm R}^N$ for the above cases 
(a) and (b), respectively. We define the submanifolds
\begin{equation}\label{Mvab}
M_v^{a,b}=\left\{q\in\Gamma_{a,b}\,\Big|\,\frac{{\mathcal H}(q)}{N}\leqslant v\right\}
\end{equation}
for the cases (a) and (b) as the subsets consisting of all points $q$ from configuration 
space with potential energies---given, in our case, by the value of the Hamiltonian 
${\mathcal H}(q)$---equal to or below a certain level $v$. As $\Gamma_a \subset \Gamma_b$, 
the relation
\begin{equation}\label{MvaMvb}
M_v^a=M_v^b \cap \Gamma_a
\end{equation}
holds which allows us to consider case (b) first and, at the end, infer the results for (a) 
by simple reasoning. Now our aim is to investigate and characterize the topology of the 
$M_v^b$ and to determine the {\em critical level(s)}\/ $v_c$ at which the topology changes 
occur. Loosely speaking, two manifolds are said to be topologically equivalent if they can 
be mapped onto each other by a smooth deformation, i.\,e., by stretching and bending, but 
not cutting or tearing. If the manifolds are not topologically equivalent, we say that a 
topology change takes place. In the following, we will explicitely characterize the topology 
and its changes for the Burkhardt model with pinning potentials 
$U(x)$, bounded below and above, with a unique infimum at $x=0$, which decrease (increase) 
monotonously for negative (positive) $x$. Without loss of generality we choose $U$ such that 
its supremum $\sup_x U(x)=0$.

\begin{figure}[hbt]
\includegraphics[width=8cm,height=9cm,clip=false,keepaspectratio=true,bb=2.8cm 21cm 9.3cm 
27.5cm]{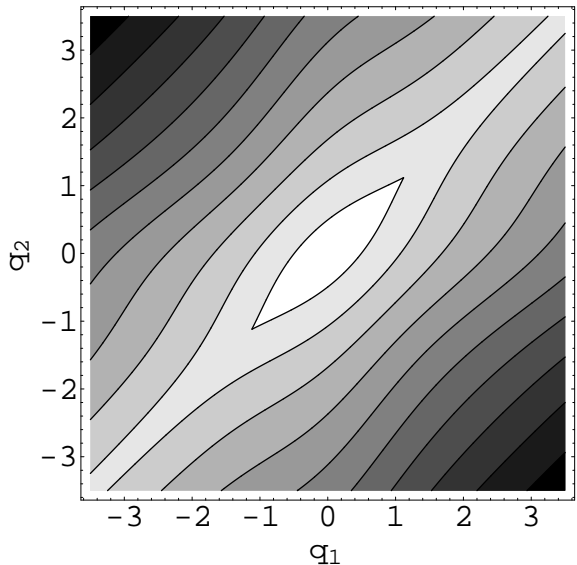}
\caption{\label{levelsets}
Submanifolds $M_v^b$ for the Burkhardt model (\ref{Hamiltonian}) with pinning potential 
$U(x)=-\ln\left(\frac{1}{2} + \frac{1}{\cosh^2 x}\right)$ \cite{PTpot} and $N=2$ degrees of 
freedom $q_1$, $q_2$. For a given level $v$, the 
submanifold $M_v^b$ consists of the area of a certain shading {\em plus all the lighter 
shaded areas}\/ (as $M_{v_1}^b \subseteq M_{v_2}^b$ for $v_1 < v_2$). For small enough $v$, 
compact $M_v^b \sim {\mathbbm I}^2$ are observed (innermost level), whereas, for larger $v$, 
non-compact manifolds $M_v^b \sim {\mathbbm R}\times{\mathbbm I}$ are found (outer levels). 
The relation $\sim$ denotes topological equivalence.
}
\end{figure}

In previous analytical calculations of topology changes in configuration space 
\cite{CaPeCo2,Angelani_etal}, {\em Morse theory}\/ was employed in order to calculate 
topological invariants. Within the standard setting of this theory as put forward originally 
by Morse \cite{Morse}, {\em compact}\/ manifolds are considered. The $M_v^{a,b}$ as defined 
in (\ref{Mvab}) are not necessarily compact, and we would have to resort to 
more sophisticated extensions of Morse theory
\cite{PalaisTerng}. The simplicity of one-dimensional models, however, may allow for a more 
direct determination of topological properties \cite{GriMo}. To this purpose, it helps the 
intuition to plot the submanifolds $M_v^b$ for the simplest 
non-trivial case of $N=2$ degrees of freedom. Figure \ref{levelsets} illustrates for an 
example of a smooth pinning potential satisfying the above boundedness and monotonicity 
conditions that, for small values of $v$, the $M_v^b$ are topologically equivalent to the 
square ${\mathbbm I}^2$ (where ${\mathbbm I}=[0,1]$ denotes the unit interval), whereas for 
$v$ above a certain critical level, topological equivalence to an infinite stripe, 
${\mathbbm 
R}\times{\mathbbm I}$, is observed.
In the following we will show that, in the general case of $N$ degrees of freedom,
\begin{equation}\label{Mvb}
M_v^b \sim \left\{
\begin{array}{c@{\quad}c@{\,\quad}c}
\emptyset & & v< U_{\inf},\\
{\mathbbm I}^N & \mbox{for} & U_{\inf} < v < 0,\\
{\mathbbm R}\times{\mathbbm I}^{N-1} & & 0 < v,
\end{array}
\right.
\end{equation}
where $\sim$ denotes topological equivalence, $\emptyset$ is the empty set, and 
$U_{\inf}=\inf_x U(x)$ is the infimum of $U$.

The simple proof goes in two steps: First, $M_v^b$ is shown to be a {\em star convex}\/ 
subset of ${\mathbbm R}^N$, i.\,e., there exists a $\widetilde{q} \in M_v^b$ such that the 
line segment from $\widetilde{q}$ to any point in $M_v^b$ is contained in $M_v^b$. This can 
be proved by showing that, for every state $q=(q_1,...,q_N)\in{\mathbbm R}^N$ and every 
$\lambda\in[0,1)$, the energy of $q'=\lambda q$ is smaller than or equal to the energy of 
$q$, i.\,e., ${\mathcal H}(q')\leqslant{\mathcal H}(q)$. The star convexity of $M_v^b$ 
implies {\em homotopical}\/ equivalence to ${\mathbbm I}^N$ (or to an $N$-ball ${\mathbbm 
B}^N$), but not necessarily topological equivalence.

In a second step, the {\em closedness}\/ of $M_v^b$ is investigated. This is done, 
analogously to the treatment in \cite{GriMo}, by studying the asymptotic behavior of the 
Hamiltonian ${\mathcal H}(\lambda q)$ in the limit $\lambda\to\infty$. Depending on the 
state $q$ considered, we find
\begin{equation}
\lim_{\lambda\to\infty} {\mathcal H}(\lambda q) = \left\{
\begin{array}{cl}
0 & \mbox{if}\quad q_i=q_j \; \forall i,j,\\
\infty & \mbox{else}.
\end{array}
\right.
\end{equation}
Hence, for negative energies, only states $q\in{\mathbbm R}^N$ with finite (Euclidean) norm 
$||q||$ are accessible, whereas states of arbitrarily large norm can be attained in the case 
of positive energies. From this observation and definition (\ref{Mvab}) it can be inferred 
that $M_v^b$ is a bounded and closed subset of ${\mathbbm R}^N$ for $v<0$, and, together 
with the star convexity shown above, it follows that $M_v^b$ is topologically equivalent to 
${\mathbbm I}^{N}$. For $v>0$, however, $M_v^b$ is unbounded and not closed. Since, for 
finite positive energies, states of arbitrarily large norm can be attained only ``in a {\em 
single}\/ spatial direction'', i.\,e., in the vicinity of the (hyper)space diagonal 
$q=\lambda(1,...,1)$, $\lambda\in{\mathbbm R}$, we conclude that $M_v^b$ is topologically 
equivalent to the product of an open interval and $N-1$ closed ones, $M_v^b\sim {\mathbbm 
R}\times {\mathbbm I}^{N-1}$. With the immediate observation that $M_v^b=\emptyset$ for 
$v<U_{\inf}$, we have accomplished a complete characterization of the topology of $M_v^b$ as 
summarized in (\ref{Mvb}).

Relation (\ref{MvaMvb}) allows to transfer this result for the case (b) of configuration 
space $\Gamma_b={\mathbbm R}^N$ straightforwardly to case (a) with $\Gamma_a=({\mathbbm 
R}_0^+)^N$. In the case of $N=2$ degrees of freedom this transfer simply consists in 
considering the positive quadrant in figure \ref{levelsets} only, giving topological 
equivalence of $M_v^a$ to ${\mathbbm I}^2$ for small values of $v$ and to ${\mathbbm R}_0^+ 
\times {\mathbbm I}$ for values above a critical level. For the general case of $N$ degrees 
of freedom, this results in a modification of (\ref{Mvb}), being of the form
\begin{equation}\label{Mva}
M_v^a \sim \left\{
\begin{array}{c@{\quad}c@{\,\quad}c}
\emptyset & & v< U_{\inf},\\
{\mathbbm I}^N & \mbox{for} & U_{\inf} < v < 0,\\
{\mathbbm R}_0^+\times{\mathbbm I}^{N-1} & & 0 < v.
\end{array}
\right.
\end{equation}

Releasing the monotonicity condition on the pinning potential $U$, additional topology 
changes should occur, but the 
essential one, being related to the localization-delocalization transition, is expected to 
persist unalteredly.

{\em Discussion of the results:} Having fully characterized the topology of the submanifolds 
$M_v^{a,b}$ for the Burkhardt model (\ref{Hamiltonian}), we found that, for our cases (a) 
and 
(b), the respective topology changes are very 
similar, although not identical, in nature. In both cases, these submanifolds are 
topologically 
equivalent to a closed $N$-ball below the transition energy (or temperature). It is only 
above the transition energy that the topology differs slightly: for case (a) with 
configuration space $\Gamma_a=({\mathbbm R}_0^+)^N$, we find equivalence to the product of 
$N-1$ closed intervals and a {\em half-open}\/ interval. For case (b) with configuration 
space $\Gamma_a={\mathbbm R}^N$, equivalence to the product of $N-1$ closed intervals and an 
{\em open}\/ interval is obtained. These topology changes are clearly not identical, but 
neither do they show a striking difference in ``strength'' like the ones observed for XY 
models with and without phase transitions \cite{CaPeCo2}.

We do not have a satisfactory definition at hand for the strength of a topology change. A 
criterion based on the variation of the Betti numbers as proposed in \cite{CaPeCo2,PeFraSpi} 
is not general enough, as the topology change reported above does not include a change of 
homotopy, and the Betti numbers remain unchanged. One might, as done in \cite{GriMo}, resort 
to the {\em surfaces} of the submanifolds $M_v$ instead of the manifolds itselves in order 
to 
obtain a change in the Betti number, but neither with this trick a significant difference in 
the strength of the topology change will be detected. Intuitively, the change in case (b) 
towards a product space including an open 
interval appears to be even {\em stronger}\/ than the change where ``only'' a half-open 
interval 
is involved; it is the latter case, however, which corresponds to a phase transition in the 
thermodynamic limit.

Our primary intention was to shed some light on the question how a sufficient criterion on 
the topology change, guaranteeing the existence of a phase transition, might look like. The 
above considerations suggest that a one-to-one correspondence between the occurrence of a 
phase transition and the strength of the underlying topology change cannot be established 
for the Burkhardt model. Hence the possibility to develop a sufficient condition, based 
exclusively on topological quantities, guaranteeing the existence of a phase transition, 
appears to be disproved by means of a counterexample for systems with non-confining 
potentials. Such a condition has, however, been proposed for a certain class of systems 
with smooth {\em confining}\/ potentials \cite{Pet_priv}, and we suspect that 
"non-confining" is the crucial property rendering an exclusively topological condition unattainable. Note that the 
analysis of the thermodynamic as well as of the topological properties of the Burkhardt model presented in this 
Letter can be straightforwardly extended to a class of systems with smooth potentials. Then, for a suitable choice of 
the on-site potentials, identical results as is (\ref{Mva}) and (\ref{Mvb}) can be obtained for systems with and 
without a phase transition, respectively. In this way it can be shown that smoothness is {\em not}\/ the crucial 
property for the attainability of a sufficient condition. Future work should try to shed some light on the 
question which ingredient, in addition to topological properties, has to be taken into 
account in order to specify a sufficient condition for the existence of a phase transition in systems with 
non-confining potentials.

In the introductory part of this Letter, the ``natural'' generalization of the concept of 
phase transitions to finite systems within the topological approach was mentioned. Observing 
that the topology change, which in the thermodynamic limit is related to the 
localization-delocalization transition of the system, is present for any finite number 
$N\geqslant 2$ of degrees of freedom, one would simply define a transition-like phenomenon 
in finite systems on the basis of this topology change and identify the critical level of 
the Hamiltonian with the transition energy. Although this might appear reasonable, it is 
clearly an unsatisfactory definition due to the lack of a sufficiency condition on the 
topology change.

{\em Summary:} The relation between thermodynamic phase transitions in classical systems and 
topology changes in their configuration space is discussed for the Burkhardt model, a 
one-dimensional solid-on-solid model. A complete characterization of the topology of the 
submanifolds $M_v$ of the configuration space is accomplished. The hypothesis---proved in 
\cite{FraPeSpi} for a certain class of systems with confining potentials---that in general a 
change of the topology within the family $M_v$ is a necessary condition in order to observe 
a phase transition, is corroborated for a larger class of systems on the basis of this 
example with non-confining potential. Considering two slightly differing versions of this 
solid-on-solid model, one showing a phase transition in the thermodynamic limit, the other 
not, we find that the difference in the ``quality'' or ``strength'' of this topology change 
appears to be insignificant. This example indicates the unattainability of a condition of 
exclusively topological nature which is sufficient as to guarantee the occurrence of a phase 
transition in systems with non-confining potentials.

\begin{acknowledgments}
Helpful comments on the manuscript from Helmut B\"uttner and Lapo Casetti
are gratefully acknowledged.
\end{acknowledgments}

\end{document}